# Effects of spin-orbit torque on the ferromagnetic and exchange spin wave modes in ferrimagnetic CoGd alloy

Boris Divinskiy,[1,*] Guanxiong Chen,[2] Sergei Urazhdin,[2] Sergej O. Demokritov,[1] and Vladislav E. Demidov[1]

[1]*Institute for Applied Physics and Center for Nonlinear Science, University of Muenster, 48149 Muenster, Germany*

[2]*Department of Physics, Emory University, Atlanta, GA 30322, USA*

We use micro-focus Brillouin light scattering spectroscopy to study the effects of spin-orbit torque on thermal spin waves in almost angular-momentum compensated ferrimagnetic CoGd alloy films. The spin-orbit torque is produced by the electric current flowing in the Pt layer adjacent to CoGd. Both the ferromagnetic and the exchange modes are detected in our measurements. The intensity and the linewidth of the ferromagnetic mode are modified by the spin-orbit torque. In contrast, the properties of the exchange mode are unaffected by the spin-orbit torque. We also find that the frequencies and the linewidths of both modes are significantly modified by Joule heating, due to the strong temperature dependence of the magnetic properties of CoGd in the vicinity of angular momentum compensation point. Our results provide insight into the mechanisms that can enable the implementation of sub-THz magnetic nano-oscillators based on ferrimagnetic materials, as well as related effects in antiferromagnets.

*Corresponding author, e-mail: b_divi01@uni-muenster.de

# I. INTRODUCTION

Recent advances in the studies of spin-orbit torques (SOTs) have opened novel opportunities for the fields of spintronics and magnonics [1-3]. In particular, SOTs have enabled the development of microwave nano-oscillators based on magnetic materials [4,5], where coherent oscillation emerges from thermally excited spin-wave modes. Following the initial demonstration [6,7], a variety of SOT-driven nano-oscillators have been proposed and experimentally realized in recent years, in efforts to improve their efficiency and coherence [8-14]. It was also theoretically shown [15, 16] and experimentally confirmed [5,6,8,17] that the mechanisms underlying the emergence of coherent dynamics can be elucidated by analyzing the evolution of the intensity and the linewidth of thermally excited modes in the sub-critical regime.

Nano-oscillators based on ferromagnetic materials operate at frequencies in the range of about 0.1 – 30 GHz [18], with the upper limit determined by the practically accessible magnitudes of static magnetic fields. On the other hand, one of the most significant challenges in modern microwave technology is the lack of compact and reliable microwave sources capable of generating signals in the frequency range 0.1 – 10 THz, which is commonly referred to as the "THz gap" [19-21]. It was recently proposed that the operational frequency of SOT oscillators based on antiferromagnetic (AFM) [22-25] and ferrimagnetic (FiM) [26] materials can be significantly higher than in ferromagnet-based oscillators, due to the large internal effective exchange fields. The latter can reach magnitudes of dozens of Tesla, enabling THz-frequency dynamics even in the absence of external magnetic fields, and paving the way for the implementation of SOT oscillators capable of filling the "THz gap".

From the point of view of technical applications, AFM materials suffer from a significant disadvantage: because of the zero net magnetic moment, excitation and detection of spin dynamics in these materials is very challenging. In contrast, in FiM materials, the antiferromagnetically coupled sublattices are not equivalent. This results in a non-zero net magnetic moment, enabling direct inductive excitation and probing of magnetization dynamics. Additionally, because of the difference in the electronic and the optical properties of the elements that constitute different sublattices, spin dynamics of these sublattices can be selectively accessed.

Among the most attractive FiM systems enabling access to the magnetization dynamics of individual sublattices are transition metal-rare earth (TM-RE) alloys [27,28]. A key benefit of these materials is that their magnetic properties can be tuned in a wide range by varying their composition and temperature [27]. In particular, because of the strong temperature dependence of the magnetization of the RE sublattice, the magnetizations of the TM and RE sublattices cancel each other at a certain composition-dependent magnetization compensation temperature $T_M$. Furthermore, the angular momenta of the two sublattices cancel at the angular momentum compensation temperature $T_A$, which is different from $T_M$ because the g-factors characterizing the TM and the RE sublattices are generally different. TM-RE also exhibit attractive electronic properties. Since the magnetism of the RE atoms is mediated by the localized $f$ electronic states with energies significantly below the Fermi level, the spin-dependent electronic transport properties of TM-RE alloys are dominated by the $d$-electrons of TM atoms. As a consequence, the spin-orbit torques act predominantly on the TM sublattice, enabling efficient SOT-driven control of the TM-RE alloys' magnetization [29].

TM-RE alloys exhibit two types of dynamical magnetic modes, as expected for FiM systems with two sublattices [30,31]. In the first mode, the magnetizations of the two sublattices remain antiparallel to each other during precession. The frequency of this mode is determined mainly by the external static magnetic field and the effective gyromagnetic ratio, and typically lies in the GHz range. These characteristics are similar to those of the dynamical modes in ferromagnets. In the second mode, called the exchange mode, the magnetizations of the two sublattices do not remain antiparallel to each other, resulting in a large contribution of exchange interaction to the dynamical mode energy. Consequently, the frequency of this mode is determined by the exchange constant, and typically falls in the THz region.

The frequencies of the ferromagnetic and the exchange modes experience strong variations at temperatures $T$ in the vicinity of the angular-momentum compensation point $T_A$. For an ideal FiM, the frequency of the ferromagnetic mode is expected to diverge at $T = T_A$ because of the divergence of the effective gyromagnetic ratio, while the frequency of the exchange mode is expected to vanish. These features are promising for the implementation of ultra-high-frequency SOT oscillators. On the one hand, very high frequency of the ferromagnetic mode can be achieved without the need for large external fields. On the other hand, the frequency of the exchange mode can be tuned down to the few-THz or sub-THz range, depending on the application requirements.

Both the ferromagnetic and the exchange modes have been experimentally observed in TM-RE alloys using ferromagnetic resonance and ultrafast optical pump-probe techniques [28,32-37]. However, the effects of SOT on these dynamical modes remain unexplored.

Here, we report an experimental study of the effects of SOT on the magnetization dynamics in the CoGd/Pt bilayer in the vicinity of the angular-momentum compensation point.

We utilize micro-focus Brillouin light scattering (BLS) spectroscopy to detect the ferromagnetic and the exchange modes, and study the dependences of their characteristics on the SOT generated by electric current in the Pt layer. By analyzing the intensity and the linewidth of these modes, we demonstrate that the effects of SOT are significant only for the ferromagnetic mode, but there is no sizable effect of SOT on the exchange mode. These observations are consistent with the general expectation that the efficiency of SOT-driven excitation is determined by the relaxation rate of the dynamical modes, which is expected to be significantly higher for the high-frequency exchange mode. We also show that the frequencies of both modes can be electronically tuned by Joule heating. The frequency of the ferromagnetic mode can reach values of up to 50 GHz at the field of 0.4 T, while the frequency of the exchange mode can be varied in the range 70-120 GHz by varying the current. Our findings are important for the practical implementation of ultra-high-frequency SOT oscillators based on FiMs, and are also likely relevant to the AFM-based SOT devices.

## II. EXPERIMENT

Figure 1(a) shows the layout of our experiment. The studied system is based on a Pt(5)/Co$_{78.1}$Gd$_{21.9}$(10) magnetic multilayer capped by Ta(3) to protect CoGd from oxidization. Here, thicknesses are in nanometers. The room-temperature saturation magnetization of the CoGd film, as determined from the vibrating-sample magnetometry measurements, is 180 kA/m. Based on this value and the experimentally determined frequencies of dynamic modes, we estimate the anisotropy constant of CoGd film to be equal to 0.08 MJ/m$^3$.

The multilayer is patterned into a square with the side of 5 μm and electrically contacted by using 120 nm thick Au electrodes. Due to the large difference in the resistivities of the CoGd,

Pt, and Ta layers (1490, 275, and 1500 nΩ*m, respectively), the electric current $I$ flowing in the plane of the multilayer is predominantly transmitted through the Pt film. The electrical current is converted by the spin-Hall effect (SHE) in Pt [38,39] into an out-of-plane spin current $I_s$. The spin current is injected into CoGd, exerting SOT on its magnetization. According to the symmetry of SHE, the effects of SOT are maximized when the static magnetic field $H_0$ is applied in plane, in the direction perpendicular to the current flow.

Since only about 16% of the total electrical current flows through the CoGd film, we assume that the contribution of SOT produced by the bulk spin-orbit interaction in the ferrimagnetic layer [40] is significantly smaller than that induced by SHE in Pt. The SHE in Ta layer plays a negligible role in the studied system because of the partial oxidation and the large resistivity of the capping Ta film. We also note that the current-induced Oersted field does not exceed 1.5 mT for the maximum current used in the experiment. Since this value is two orders of magnitude smaller than the strength of the static magnetic field (0.1 – 0.4 T), we assume negligible effects of the Oersted field.

We characterize the effects of the driving current on the dynamical modes by using micro-focus BLS spectroscopy [41]. The probing laser light with the wavelength of 532 nm is focused into a diffraction-limited spot on the surface of the CoGd film, and the spectrum of light inelastically scattered from the dynamical magnetization is analyzed. The incident beam intensity of about 0.1 mW is sufficiently low to ensure that the perturbation of the magnetic system by the probing light is negligible. The high sensitivity of BLS enables detection of thermally excited spin-wave modes (magnetic fluctuations), which are always present at nonzero temperatures even in the absence of the driving electric current, allowing the characterization of the magnetic system in the subcritical regime.

## III. RESULTS AND DISCUSSION

Figure 1(b) shows a representative BLS spectrum of magnetic fluctuations recorded at the magnetic field $\mu_0 H_0 = 0.1$ T, at room temperature $T_0 = 295$ K. The spectrum exhibits a well-defined peak with a pronounced shoulder on its high-frequency tail, indicating the existence of two dynamic modes in the studied system. The spectrum is well-approximated by a sum of two Lorentzian functions, enabling accurate determination of the central frequencies of the two modes. We will refer to them as the low-frequency (LF) and the high-frequency (HF) mode. As the external field $\mu_0 H_0$ is increased from 0.1 T to 0.4 T, the central frequency of the LF mode monotonically increases by a factor of 1.5 from 28 to 42 GHz, while the frequency of the HF mode remains nearly constant at 65 GHz, as shown in Fig. 1(c). Note that at $\mu_0 H_0 > 0.3$ T, the peak corresponding to the LF mode strongly overlaps with that of the HF mode, so the latter becomes difficult to distinguish in the measured spectra. The obtained dependences agree well with the theory of magnetization dynamics in FiMs [30] and previous experimental observations [34], allowing us to identify the LF and the HF mode as the ferromagnetic and the exchange mode, respectively. Indeed, the frequency of the ferromagnetic mode is expected to increase with the increase of $H_0$, while the frequency of the exchange mode is expected to be nearly independent of $H_0$, since it is determined mainly by the effective exchange field.

Next, we study the effects of the electric current on the characteristics of the observed modes. These effects generally include variations of the intensity of fluctuations and of the effective damping [17], resulting in the variations of the intensity and the linewidth of the spectral peaks, respectively. For the direction of $H_0$ shown in Fig. 1(a), SOT induced by the positive current $I$, as defined in this Figure, is expected to enhance magnetic fluctuations and decrease the effective mode damping. The opposite effects are expected for negative current.

Figure 2(a) shows the current dependence of the integral intensity $E$ of the measured BLS spectra. This dependence is clearly asymmetric with respect to the current direction, even though it is dominated by the symmetric quadratic contribution (dashed curve in Fig. 2(a)) that can be attributed to Joule heating. The asymmetric deviations from the quadratic dependence become particularly pronounced at large currents $|I|>10$ mA. Figures 2(b) and 2(c) show the current dependences of the integral intensities $E_f$, $E_{ex}$ of the peaks associated with the ferromagnetic and the exchange mode, respectively, obtained from the Lorentzian fits of the measured spectra similar to that shown in Fig. 1(b). To highlight the effects of SOT, which depend on the direction of the current, the data for $I>0$ and for $I<0$ are shown on the same plot as a function of the current magnitude.

Figure 2(b) clearly demonstrates that for the ferromagnetic mode, the integral intensity is generally larger at $I>0$ than at the same magnitude of $I<0$. This result is consistent with the effects of SOT, which are expected to enhance magnetic fluctuations at $I>0$, and suppress them at $I<0$ [17]. We note that the asymmetry between the opposite current directions is strongly nonlinear. In particular, the intensity at $I=12$ mA is only 7% larger than at $I=-12$ mA, while at the maximum applied current $I_{max}=17$ mA, the asymmetry defined as $2(E_f(+I_{max}) - E_f(-I_{max}))/(E_f(+I_{max})+E_f(-I_{max}))$ reaches about 30%. According to the general theory of spin torque, the current dependence of intensity associated with the SOT can be described by $E=E_0(1-I/I_C)^{-1}$ [15]. Here, $I_C$ is the critical current, at which the natural damping is expected to become completely compensated by SOT. This dependence is valid for ferromagnetic materials only in the limit of negligible Joule heating and current-independent mode frequency [15], which cannot quantitatively account for our results. Nevertheless, using this dependence as an approximation, we can estimate the value $I_C \approx 50\text{-}100$ mA of the critical current for the studied system. This value corresponds to the average current density of $10^{12}$ A/m$^2$ in the bilayer, which is comparable to the values typical for the previously demonstrated SOT oscillators based on ferromagnetic metals [4,5].

In contrast to the ferromagnetic mode, the results for the exchange mode do not indicate any sizable effects of SOT. The measured integral intensity $E_{ex}$ of the peak increases with current, but this increase is independent of the current direction, within the experimental error (Fig. 2(c)). This result is not surprising, since $I_C$ is proportional to the relaxation rate [15], which in the Gilbert damping approximation is proportional to the mode frequency. Since the frequency of the exchange mode is significantly higher than that of the ferromagnetic mode, the corresponding critical current is expected to be much larger. One can estimate that, at $I_{max}=17$ mA, the intensity of the exchange mode is expected to be enhanced by no more than a few percent, consistent with the data of Fig. 2(c).

SOT also modifies the mode relaxation rate, which is expected to be manifested by the asymmetric current dependence of the peak linewidth. The current dependences of the spectral widths of the peaks are shown in Figs. 3(a) and 3(b) for the ferromagnetic and the exchange mode, respectively. Similarly to the intensities, the effect of SOT on the linewidth is sizeable only for the ferromagnetic mode. At $I>0$ (point-up triangles in Fig. 3(a)), the linewidth is generally smaller than at $I<0$ (point-down triangles). In contrast, for the exchange mode (Fig. 3(b)), the differences between the linewidths for the opposite current directions are within the experimental uncertainty.

The data shown in Figs. 2 and 3 clearly demonstrate that, at $I>0$, SOT acts as the anti-damping torque, increasing the intensity of thermal fluctuations and decreasing the spectral linewidth. Furthermore, the relation between the effects of SOT and the frequency of the specific mode is consistent with the dependences previously established for ferromagnets. Thus, while FiMs can provide high oscillation frequencies at moderate static magnetic fields, complete compensation of the natural damping, needed for the excitation of magnetization auto-oscillations at these frequencies, is expected to require very large driving currents. Thus, practical implementation of near-THz SOT oscillators utilizing exchange mode is a challenging task that may require technological and/or scientific breakthroughs in the efficiency and selectivity of SOT-driven mode excitation. The latter may be accomplished by taking advantage of the energy and momentum selection rules involved in the excitation of magnetization dynamics by spin injection [42].

We note that the effects of Joule heating clearly play a significant role in the observed behaviors of the dynamical modes. These effects are expected to be particularly large in the vicinity of the angular-momentum compensation point of FiM, where the frequencies of both the ferromagnetic and the exchange mode are expected to rapidly vary with temperature. Figure 4(a) shows the dependence of the mode frequencies on the experimental temperature in the absence of current, confirming this expectation. As the temperature is increased above 295 K, the frequency of the ferromagnetic mode first increases, reaches a maximum of about 50 GHz at $T = 304$ K, and then monotonically decreases at higher $T$. The frequency of the exchange mode can be reliably determined at $T > 315$ K, where it monotonically increases from 76 to 114 GHz with increasing $T$.

These temperature dependences agree with the previous theoretical predictions and reported observations for the ferromagnetic and exchange modes in FiMs [32,33,43]. According to the established models, the frequency of the exchange mode reaches a minimum, while that of the ferromagnetic mode reaches a maximum at the angular momentum compensation temperature $T_A$. Thus, the data of Fig. 4(a) indicate that, for the studied system, the angular momentum compensation temperature is $T_A = 304$ K, 9 K above the room temperature $T_0 = 295$ K. We note that, in CoGd alloys, the magnetization compensation temperature $T_M$ is typically about 30 – 60 K smaller than $T_A$ [33,44]. We conclude that, for our samples, the magnetization compensation temperature $T_M$ is below the room temperature. Therefore, the net magnetic moment of the studied films is determined by the Co sublattice within the entire temperature range used in the experiments.

The effects of Joule heating on the frequencies of the two modes are illustrated in Fig. 4(b). The frequency of the ferromagnetic mode increases with increasing current, reaches a maximum of about 50 GHz at $|I| = 8$ mA, and monotonically decreases at $|I| > 8$ mA. The exchange mode becomes distinguishable in the spectrum at $|I| > 12$ mA, where its frequency monotonically increases with increasing current magnitude. Note that these variations are nearly identical for the two opposite directions of current, confirming that they are dominated by Joule heating.

By comparing Figs. 4(a) and 4(b), we conclude that at $T_0=295$ K the current-dependent temperature of the sample becomes equal to the angular-momentum compensation temperature $T_A= 304$ K at $|I| = 8$ mA. At $T>T_A$, the frequency of the ferromagnetic mode rapidly decreases, while that of the exchange mode decreases with increasing temperature (Fig.4(a)). This explains the rapid variations of the mode linewidths at large current magnitudes, observed for both current directions (Fig. 3). Indeed, in the Gilbert approximation, the increase (decrease) of the mode frequency is expected to result in the increase (decrease) of its linewidth. Accordingly, the linewidth of the ferromagnetic mode rapidly decreases with increasing current magnitude, while that of the exchange mode increases.

## IV. CONCLUSIONS

In conclusion, we have experimentally studied the dynamic magnetic modes in an almost compensated ferrimagnetic CoGd film, and their controllability by SOT induced by electrical current in the adjacent Pt layer. Our results indicate that CoGd may be suitable for the implementation of SOT oscillators with frequencies of up to several tens of GHz, achievable at moderate static magnetic fields. Our data also indicate that SOT oscillators based on the ferrimagnetic exchange mode may allow one to achieve frequencies approaching the THz range, but their operation would likely require extreme driving current densities that may be challenging to achieve in real devices. We expect similar constraints to be also relevant to antiferromagnet-based devices. These findings provide an important insight into the practical aspects of the development of ultra-high-frequency SOT-driven devices for spintronic and magnonic applications.

This work was supported in part by the Deutsche Forschungsgemeinschaft (Project No. 423113162) and the NSF award ECCS-1804198.

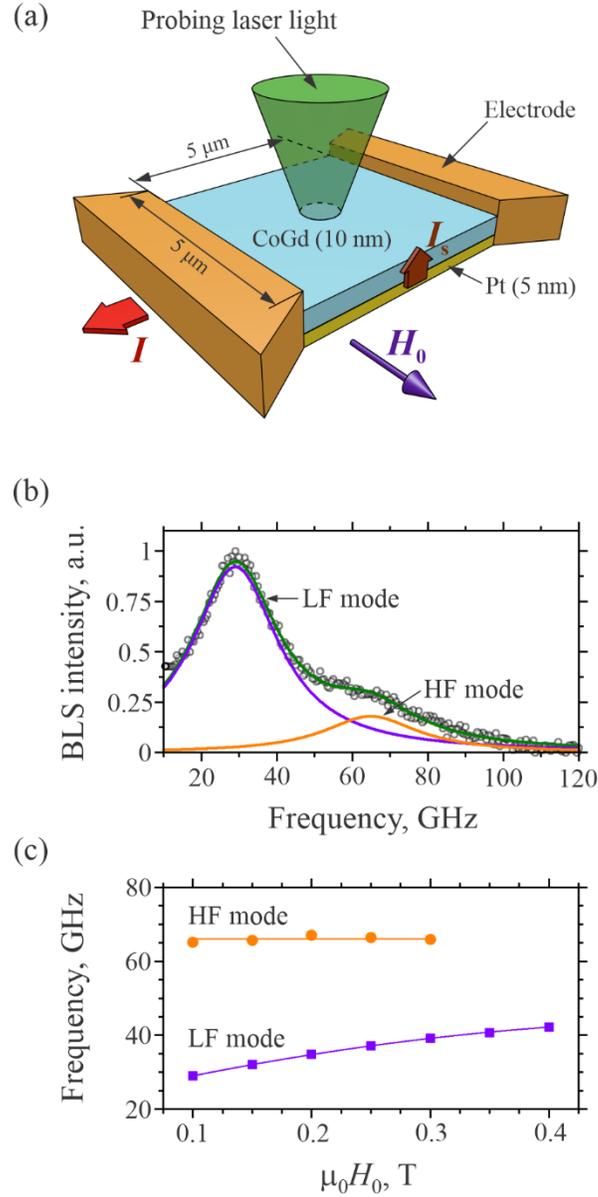

**Figure 1** (a) Schematic of the experiment. (b) Representative BLS spectrum measured at $\mu_0 H_0 =$ 0.1 T. Symbols are experimental data. Curves are the Lorentzian fits for the low-frequency (LF) and the high-frequency (HF) modes, and their sum. (c) Field dependences of the central frequencies of the LF and HF modes. Symbols are the experimental data, curves are guides for the eye. All the data were obtained at room temperature $T_0 = 295$ K.

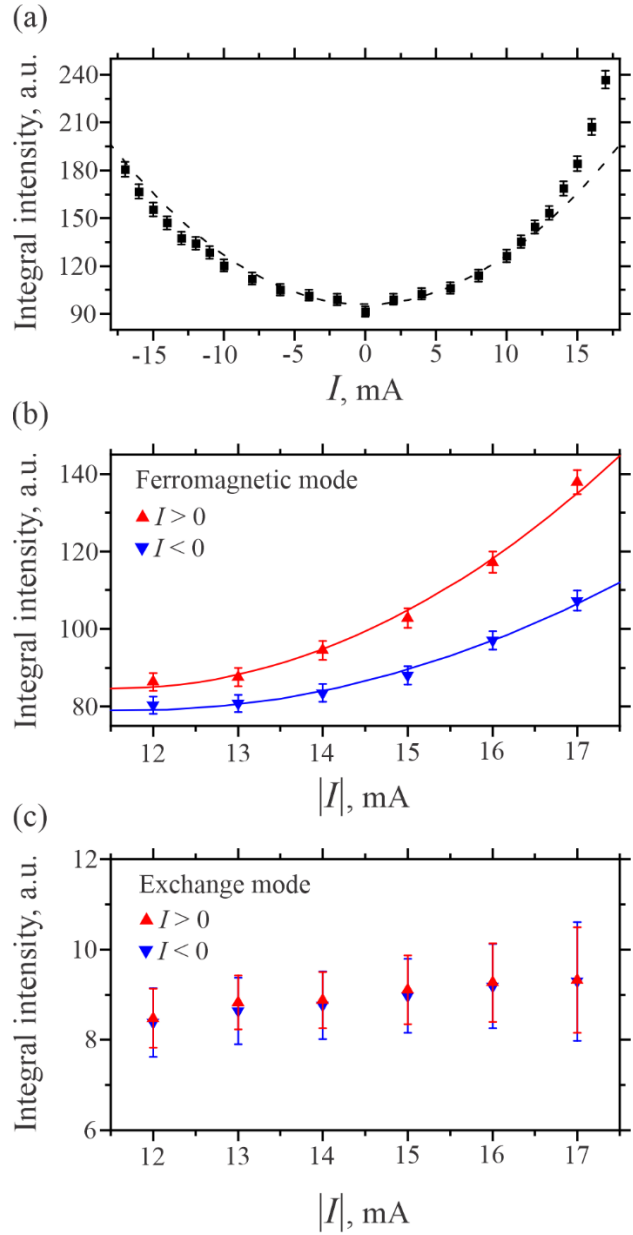

**Figure 2** (a) Current dependence of the total integral intensity of the measured BLS spectra. (b), (c) integral intensity for the ferromagnetic (b) and the exchange (c) modes, obtained from the Lorentzian fits of the corresponding spectral peaks. Symbols are the experimental data. Dashed curve in (a) shows the result of a quadratic fit. Curves in (b) are guides for the eye. Error bars show the uncertainty of the data. The data were recorded at $\mu_0 H_0 = 0.4$ T.

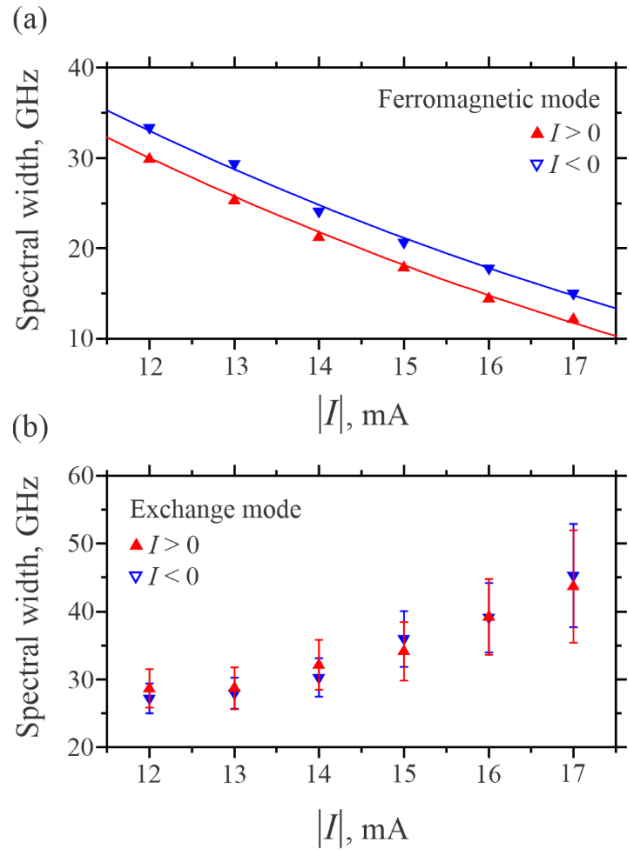

**Figure 3** Current dependences of the spectral linewidth of the peaks corresponding to the ferromagnetic (a) and the exchange (b) mode. Symbols are the experimental data, curves are guides for the eye. Error bars show the fitting uncertainty. For clarity, the error bars are shown only if the error exceeds the size of the symbols. The data were recorded at $\mu_0 H_0 = 0.4$ T.

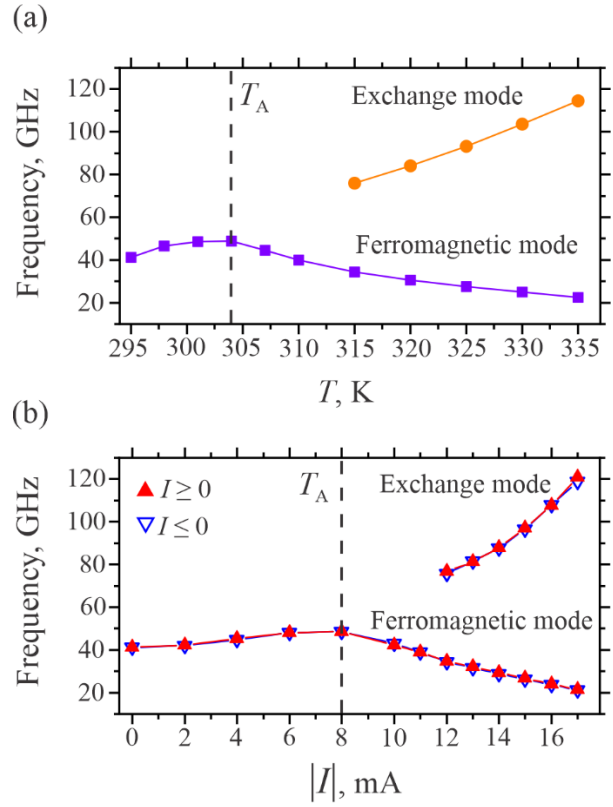

**Figure 4** Temperature (a) and current (b) dependences of the central frequencies of the ferromagnetic and the exchange modes, as labeled. $T_A$ marks the angular-momentum compensation temperature. Point-up and point-down triangles in (b) show the data for the positive and the negative currents, respectively. Symbols are the experimental data, curves are guides for the eye. The data were recorded at $\mu_0 H_0 = 0.4$ T.